\documentclass{PoS}

\newcommand{\etal}{{\sl et al. }}
\newcommand{\degs}{$^{\circ}$}
\newcommand{\ergs}{$\,$erg$\,$s$^{-1}$}
\newcommand{\eg}{{\sl e.g. }}
\newcommand{\Chandra}{{\it Chandra }}
\newcommand{\arcsec}{$''$}

\title{Determining the nature of the faint X-ray source population near the Galactic Centre}

\ShortTitle{Faint X-ray sources near the Galactic Centre}

\author{\speaker{Reba M. Bandyopadhyay}\\
        Dept. of Astronomy, University of Florida, Gainesville, FL 32611 USA\\
        E-mail: \email{reba@astro.ufl.edu}}

\author{Andrew J. Gosling, Katherine M. Blundell, Philipp Podsiadlowski\\
        Dept. of Astrophysics, University of Oxford, Oxford OX1 3RH, UK}

\author{Stephen E. Eikenberry, Valerie J. Mikles\\
        Dept. of Astronomy, University of Florida, Gainesville, FL 32611 USA}

\author{James C.A. Miller-Jones\\
        Astronomical Institute ``Anton Pannekoek'', University of Amsterdam, 1098 SJ Amsterdam, NL}

\author{Franz E. Bauer\\
        Columbia University, New York, NY 10027, USA}

\abstract{We present results of a multi-wavelength program to study
the faint discrete X-ray source population discovered by Chandra in
the Galactic Centre (GC).  From IR imaging obtained with the VLT we
identify candidate K-band counterparts to 75\% of the X-ray sources in
our sample.  By combining follow-up VLT $K$-band spectroscopy of a
subset of these candidate counterparts with the magnitude limits of
our photometric survey, we suggest that only a small percentage of the
sources are HMXBs, while the majority are likely to be canonical LMXBs
and CVs at the distance of the GC.  In addition, we present our
discovery of highly structured small-scale (5-15\arcsec) extinction
towards the Galactic Centre.  This is the finest-scale extinction
study of the Galactic Centre to date. Finally, from these VLT
observations we are able to place constraints on the stellar
counterpart to the ``bursting pulsar'' GRO~J1744-28.}

\FullConference{VI Microquasar Workshop: Microquasars and Beyond\\
		 September 18-22, 2006\\
		 Como, Italy}

\begin{document}

\section{The {\it Chandra} Galactic Centre Survey}
\label{sec:intro}

\begin{figure*}[!t]
\centering{
  \includegraphics[width=.9\textwidth]{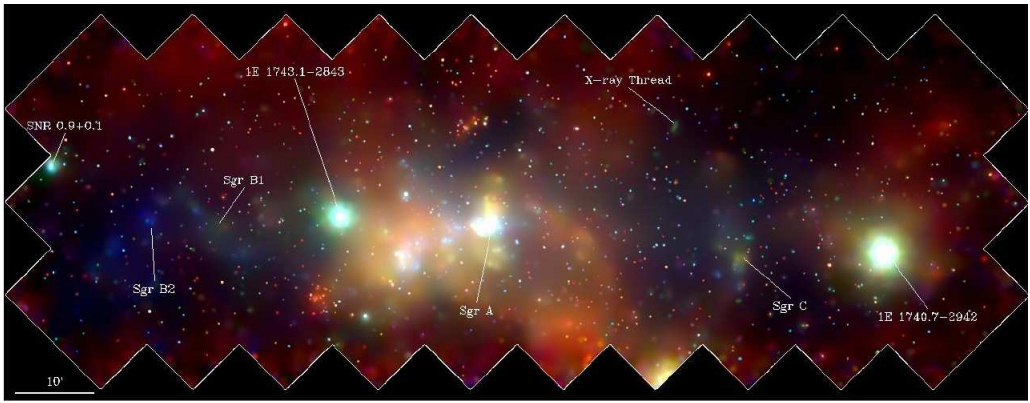}}
 \caption{Full \Chandra ACIS-I mosaic of the Galactic Centre (Wang \etal 2002).}
\label{fig:x-rayfield}
\end{figure*}

In July 2001 Wang \etal (2002) performed an imaging survey with {\it
Chandra}/ACIS-I of the central 0.8$\times$2\degs of the Galactic
Centre (GC), revealing a large population of previously undiscovered
discrete weak sources with X-ray luminosities of $10^{32}-10^{35}$\ergs 
(Figure 1).  The nature of these $\sim$800 newly detected
sources, which may contribute $\sim$10\% of the total X-ray emission of
the GC, is as yet unknown.  In contrast to the populations of faint
AGN discovered from recent deep X-ray imaging out of the Galactic
plane, our calculations suggest that the extragalactic contribution to
the hard point source population over the entire Wang \etal survey is
$\leq$ 10\%, consistent with the log(N)-log(S) function derived from
the {\it Chandra} Deep Field data (\eg Brandt \etal 2001).  The harder
($\geq$3 keV) X-ray sources (for which the softer X-rays have been
absorbed by the interstellar medium) are likely to be at the distance
of the GC, while the softer sources are likely to be foreground X-ray
active stars or cataclysmic variables (CVs) within a few kpc of the
Sun.  The distribution of X-ray colours suggests that only
a small fraction of the {\it Chandra} sources are foreground objects.
The combined spectrum of the discrete sources shows emission lines
characteristic of accreting systems such as CVs and X-ray binaries
(XRBs).  These hard, weak X-ray sources in the GC are therefore most
likely a population of XRBs; candidate classes include quiescent black
hole binaries or quiescent low-mass XRBs, CVs, and high-mass
wind-accreting neutron star binaries (WNSs).

\begin{figure*}[!t]
\centering{
  \includegraphics[width=.75\textwidth]{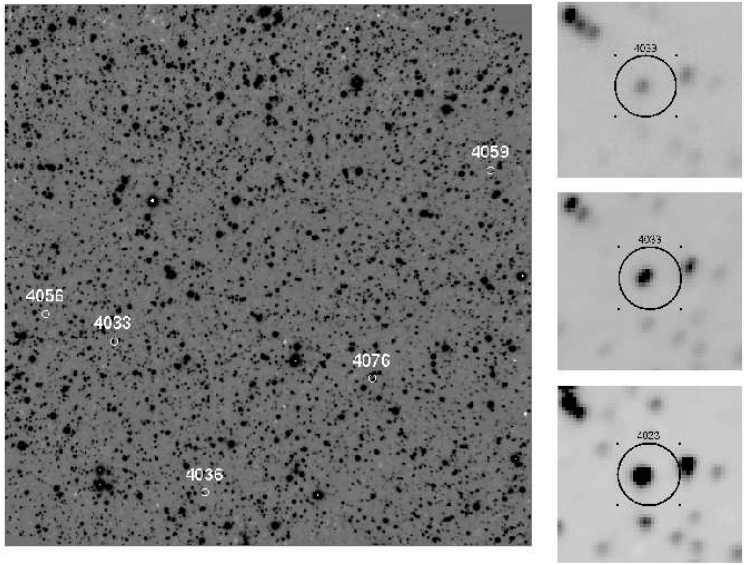}}
 \caption{Left: an example ISAAC $K_{s}$-band field (2.5
  arcmin$^{2}$), showing the positions of 5 \Chandra X-ray sources.
  Right: Zoom-in view (8\arcsec $\times$ 8\arcsec) of the 1.3\arcsec\
  error circle of one of the \Chandra sources in this field, overlaid
  on the $J$ (top), $H$ (middle), and $K_{s}$ (bottom) ISAAC images.}
\label{fig:IRfield}
\end{figure*}

\section{VLT Observations}
\label{sec:obs}

The first step in determining the nature of this population is to
identify counterparts to the X-ray sources.  Achievement of our goals
requires astrometric accuracy and high angular resolution to overcome
the confusion limit of the crowded GC.  We used ISAAC on the VLT to
obtain high-resolution $JHK$ images in order to identify a
statistically significant number of counterparts to the X-ray sources
on the basis of the {\it Chandra} astrometry.  We imaged 26 fields
within the {\it Chandra} survey region, containing a total of 79 X-ray
sources.  The average extinction towards the GC is $K\sim$2--3.
Therefore with our images, which have a magnitude limit of $K$=20, we
will detect any XRBs with {\it early-type} (O, B, A) or {\it evolved}
mass donors, all of which would have intrinsic $K$ magnitudes $\leq$17
at 8.5 kpc.

After analyzing the photometric data (Bandyopadhyay \etal 2005), we
selected 28 candidate counterparts (magnitudes $K\sim$12-17) for
follow-up $K$-band spectroscopy, to search for the characteristic
accretion signatures, such as Brackett $\gamma$ emission, that would
denote the identity of true X-ray source counterparts.  The long slit
spectra were obtained with ISAAC in service mode between
April-September 2005, using a 1\arcsec slit width ($R=$450).

\section{Results: Imaging}
\label{sec:phot}

\begin{figure*}[!t]
\vspace{-0.5cm}
\hspace{-1cm}
\rotatebox{-90}{
\includegraphics[width=0.43\textwidth]{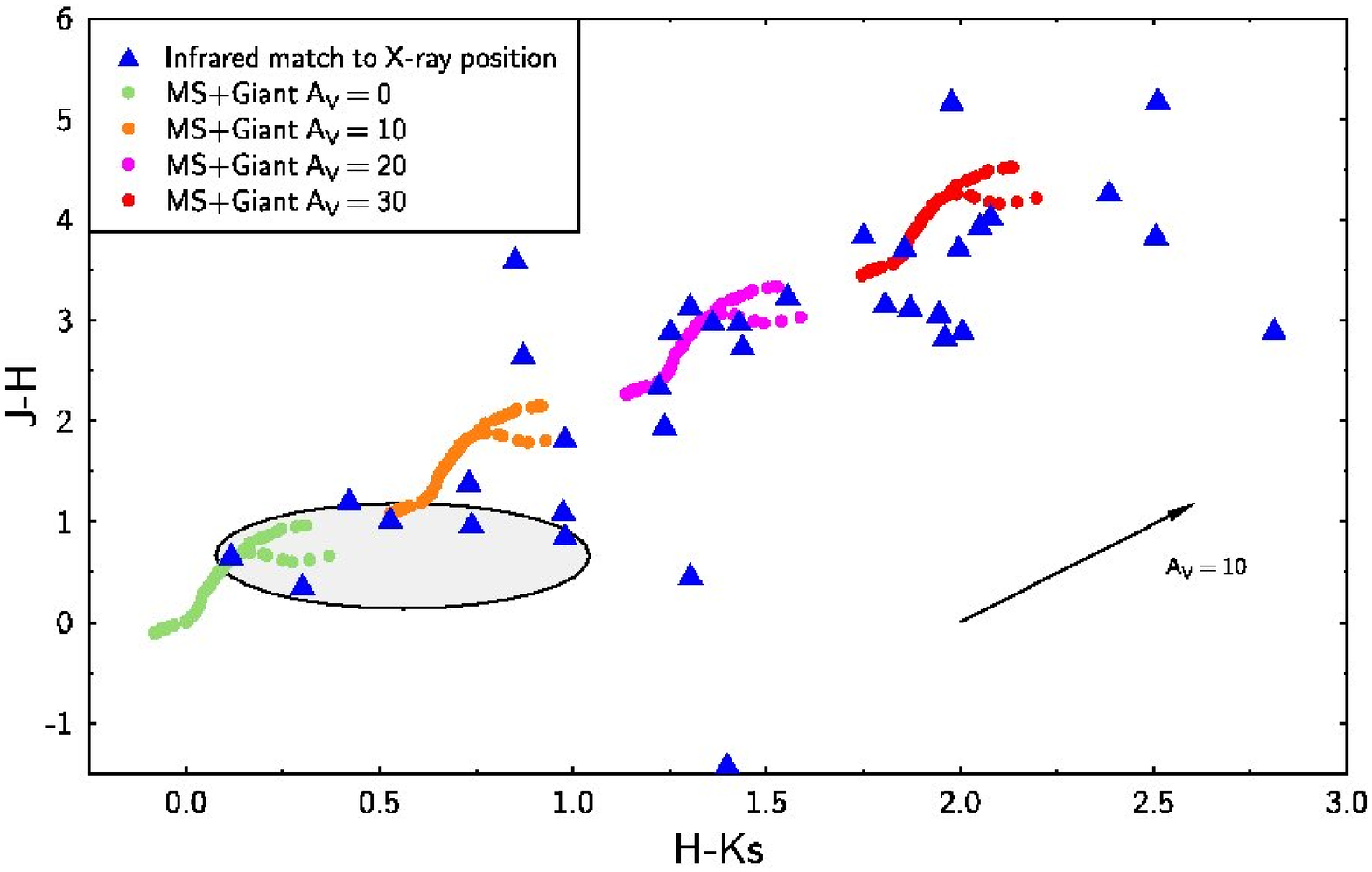}}
\hspace{-1.4cm}
\rotatebox{-90}{
\includegraphics[width=0.43\textwidth]{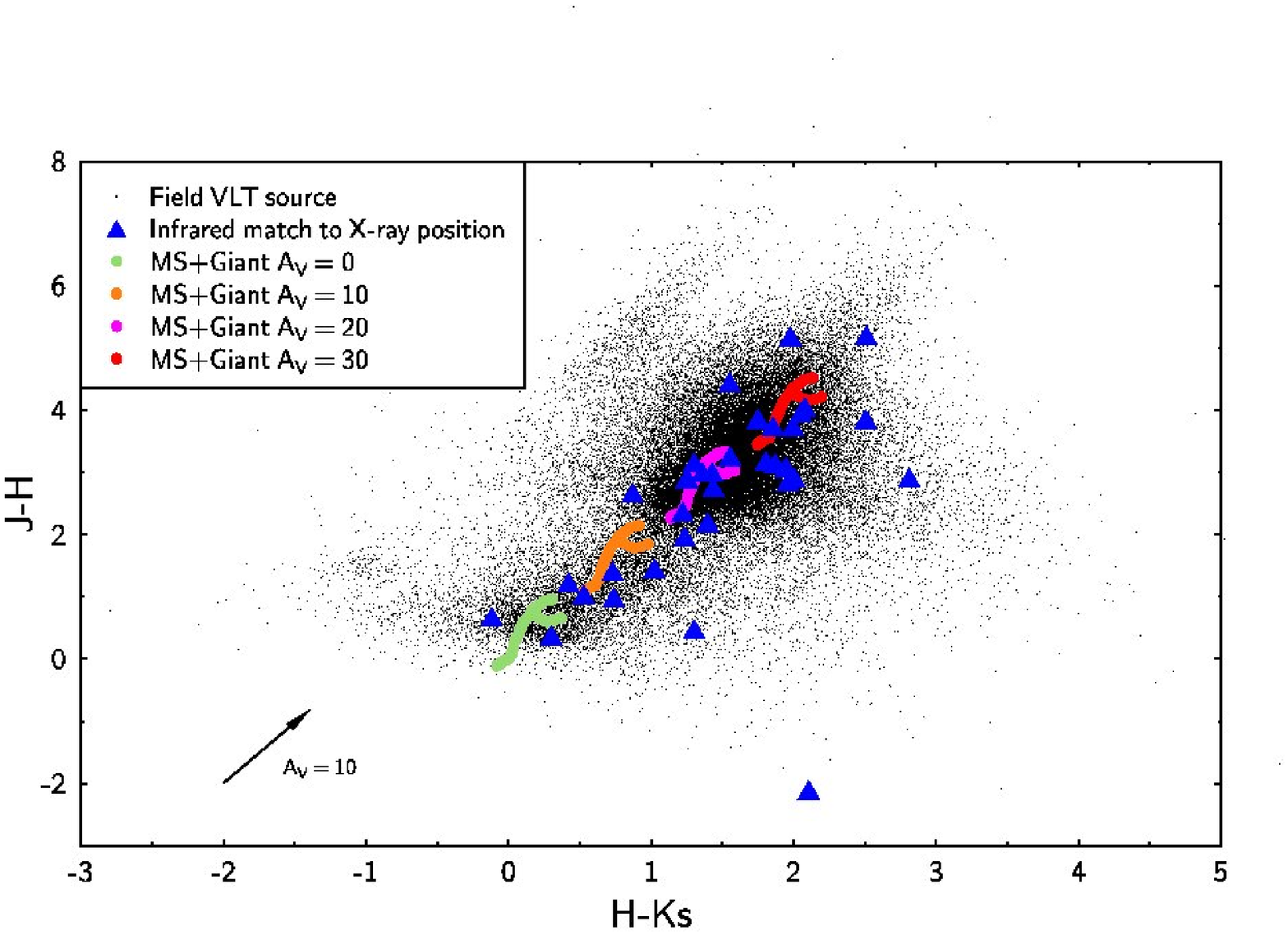}}
\vspace{-1cm}
\caption{{\em Left:} Colour-colour diagram of astrometrically-selected
  candidate IR counterparts to the \Chandra sources.  The shaded oval
  indicates where unreddened AGN/QSOs would most likely be located.
  The theoretical main sequence and giant branches are indicated at
  visual extinctions of 0, 10, 20, 30 ($A_{K}/A_{V} =$ 0.11).  {\em
  Right:} Colour-colour diagram showing all sources in our VLT fields.
  This illustrates that the vast majority of field stars are
  consistent with highly reddened stars; thus most of the stars
  (including potential X-ray counterparts) are at the distance of the
  GC (or beyond).}
\label{colour_colour}
\end{figure*}

For 65\% of the X-ray sources in our VLT fields, there are 1-2
resolved $K$-band sources within the 1\arcsec {\it Chandra} error circle
(Figure 2); only a small number of X-ray sources have more than two
potential counterparts.  Over 50\% of the {\it Chandra} sources have
no potential $J$-band counterparts, and only a few of the potential IR
counterparts have colours consistent with unreddened foreground stars
(Figure 3).  This is consistent with the expectation that the majority
of the detected X-ray sources are heavily absorbed and thus are at or
beyond the GC.

The magnitude and colour distribution of the identified candidate
counterparts is redder than expected for WNS systems.  For an average
GC extinction of $A_{K}\sim$2-3, the peak of the expected reddened $K$
magnitude distribution for the WNSs is $\sim$16-17.  The peak of the
observed reddened $K$ magnitudes for the potential counterparts is
$\sim$14-15, with an {\it (H-K)} colour of $\sim$1-2, as expected for
later-type stars.  However, some potential counterparts do have
colours consistent with early-type stars.

\subsection{Extinction}

\begin{figure}
\centering{
\includegraphics[width=0.9\columnwidth]{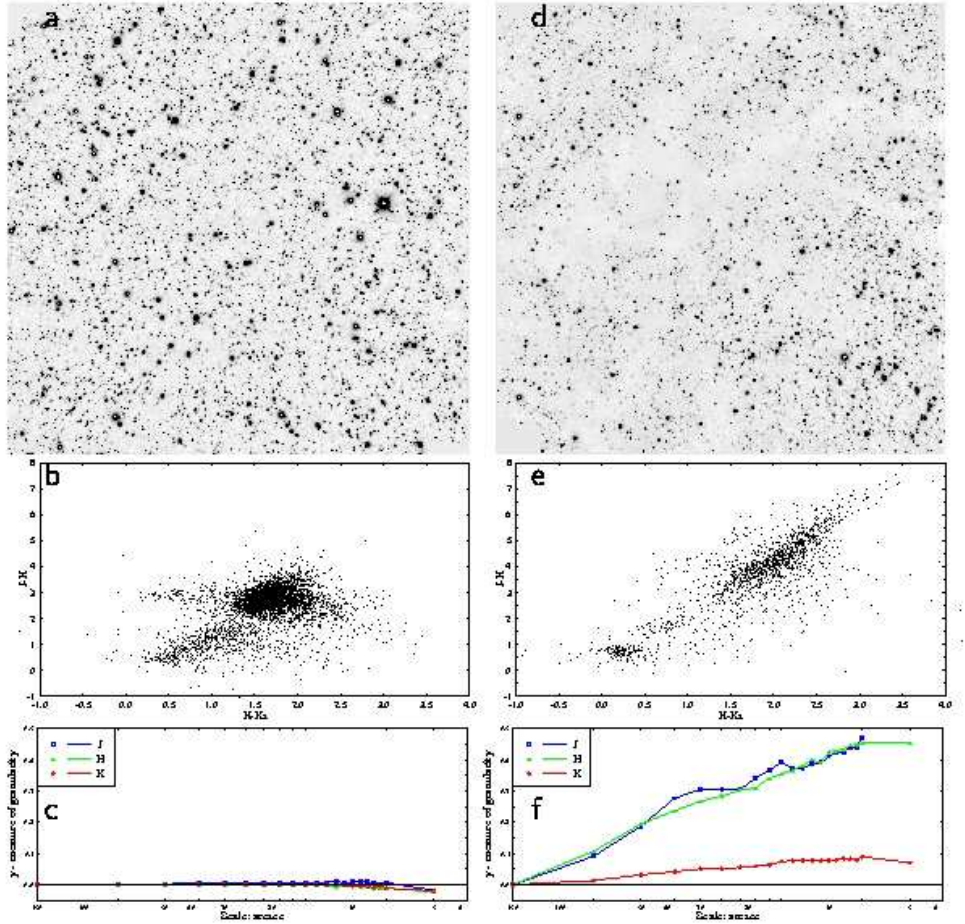}}
\caption{(a) $K_{s}$-band image of a field with no apparent structure
in the stellar distribution. (b) Colour-colour diagram of the stars in
the field shown in (a).  There are two main loci of stars: the local
population to the bottom left, and the Galactic centre population in
the centre. (c) The measures of granularity for the field shown in
(a). In all three bands it does not deviate from zero, as expected for
a random distribution. (d) $K_{s}$-band image of a field with obvious
regions of low stellar density compared to the field average. (e)
Colour-colour diagram for the field shown in (d). Note, compared to
the colour-colour diagram shown in (b), the locus for the GC stars is
extended to high reddening. (f) The measure of granularity for the
field shown in (d) shows that there is measurable structure in all
three bands.}
\end{figure}

There are no $K$-band counterparts for $\sim$35\% of the {\it Chandra}
sources.  This is larger than the expected fraction of background AGN
from the CDF estimate, though other groups have predicted larger
fractions (up to 50\%).  However, we find that the extinction in the
GC is highly structured on scales as small as 5\arcsec--15\arcsec
(corresponding to a physical scale of 0.2--0.6 pc at 8.5 kpc), even in
the $K$-band (Gosling \etal 2006).  Although the average extinction in
the GC is $A_{K}\sim2.3$, in some of the observed dust patches and
lanes, this extinction can rise to levels of as much as $A_{K}\sim6$.

Comparison of areas of low apparent stellar density with the
colour-colour diagrams of the fields confirms that this
``granularity'' is due to extinction rather than to any intrinsic
``clumpiness'' in the underlying stellar distribution (Figure 4).
These results are in agreement Schultheis \etal (1999), who suggested
that structures in the GC stellar population distribution smaller than
the resolution of their imaging ($\sim$1\arcsec) were responsible for
observed double peaks in histograms of star counts versus $A_{V}$ in
the GC.  Our results are also consistent with the finding of Nishiyama
\etal (2006) that the IR extinction varies between different lines of
sight in regions around the GC; thus the previously assumed
universality of the IR extinction law for the GC is not valid,
especially on small angular scales.  Therefore we need to carefully
determine which X-ray sources actually have no IR counterpart down to
the $K$=20 magnitude limit, and which are located in areas of locally
heavy extinction.

\section{Results: Spectroscopy}
\label{sec:spec}

\begin{figure}[!t]
\centering{
  \includegraphics[width=0.7\columnwidth]{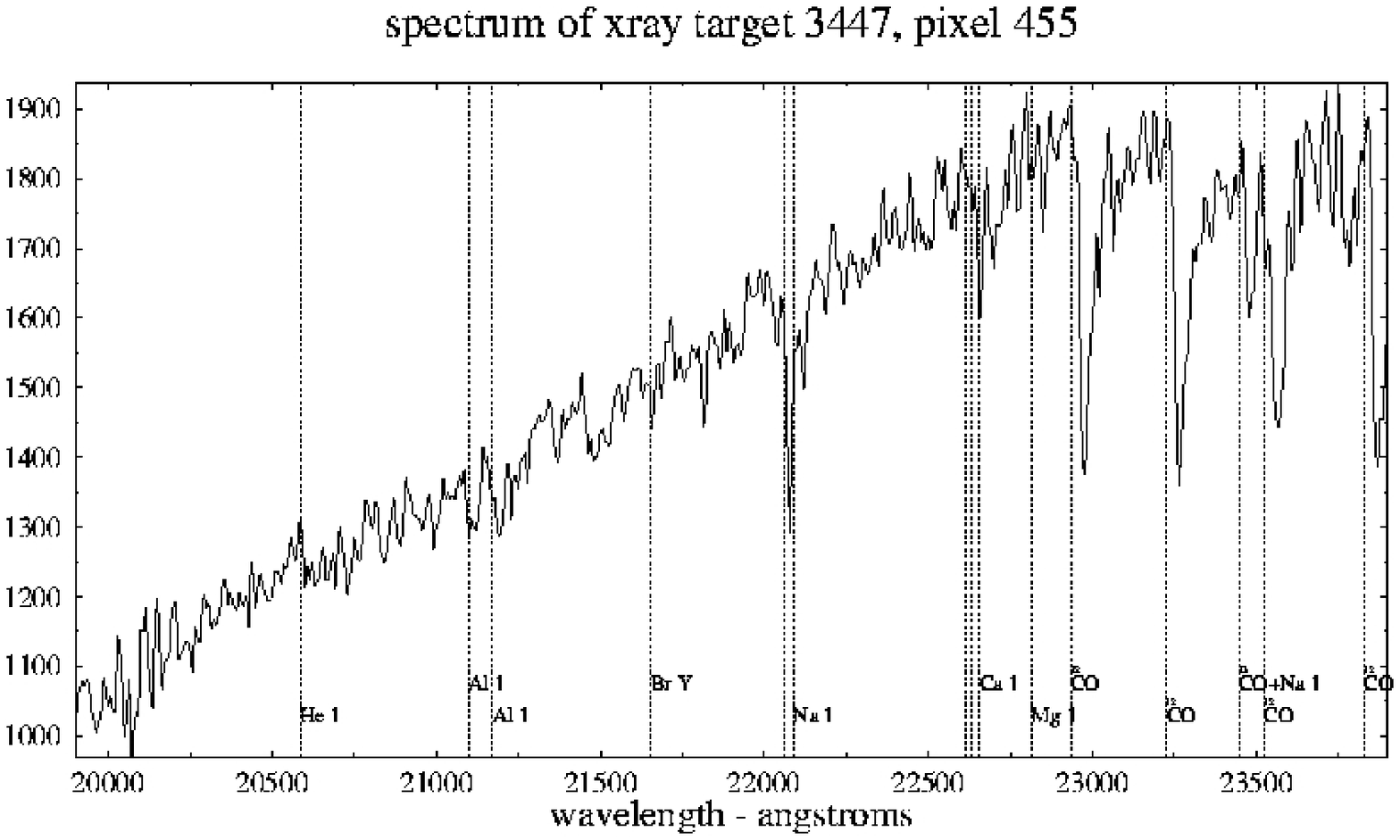}}
\vspace{-0.5cm}
  \caption{Example of one of the $K$-band spectra obtained of possible
  counterparts to the GC X-ray sources.  From the sample of 36 targets
  for which spectra were obtained with ISAAC, none exhibited emission
  lines characteristic of accreting binaries.  The spectra of the
  majority of the stars observed appeared to be K or M giants, with
  strong CO bandheads as well as other metal absorption lines.}
\end{figure}

The primary accretion signature in the $K$-band which distinguishes a
true X-ray counterpart from a field star is strong Brackett $\gamma$
emission; this technique of identifying XRB counterparts has been
verified with observations of several well-studied GC XRBs (see \eg
Bandyopadhyay \etal 1999).  As these \Chandra sources are weaker
in X-rays than the previously known population of Galactic XRBs, and
thus have lower accretion rates, the emission signature will likely be
somewhat weaker than in the more luminous XRB population.  

The spectra indicate that almost all of the candidate counterparts are
K/M giant stars, identified by the strong CO absorption bands above
2$\mu$m and several metal (Ca, Na, Mg) absorption lines (Figure 5).
None of the observed spectra exhibited the emission line signatures
characteristic of accreting binaries.  A possible explanation for this
result is that the accretion signatures could be too weak to be
measureable, for example if the accretion rate was low at the time of
observation, or if the emission was self-absorbed by the mass donor
(as was observed for the transient V404 Cyg in quiescence; Shahbaz
\etal 1996).  However, the Br $\gamma$ accretion signature is clearly
detected in the IR spectra of CVs, which are only weak X-ray emitters
with a similar X-ray luminosity range to the \Chandra sources
(Dhillon \etal 1997).


A more likely explanation for the lack of observed emission in the
candidate spectra is that the stars we observed are not the true
counterparts to the X-ray sources.  Our imaging survey had a limiting
magnitude of $K$=20, so our VLT survey would detect XRBs with either
early-type or evolved mass donors.  Therefore it is likely that the
majority of the true IR counterparts belong to a lower mass population
of stars that at GC distances are fainter than the limits of our
survey.  

\subsection{The Bursting Pulsar}

\begin{figure}
\centering{
\includegraphics[width=0.6\columnwidth]{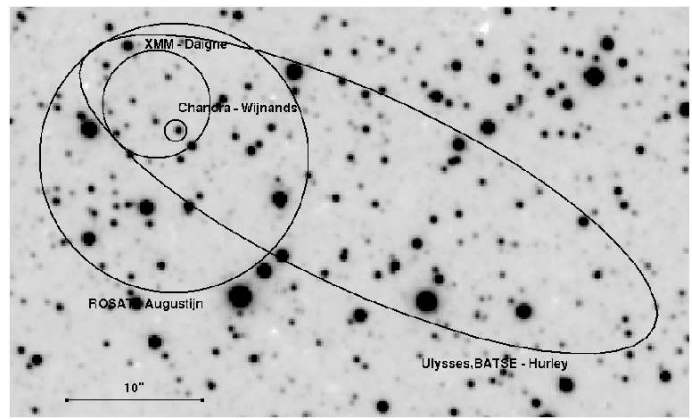}}
\caption{Position error circles for GRO~J1744-28 as determined by different
X-ray instruments. These are overlaid on the $K_{S}$-band image taken with
the VLT.}
\label{finder}
\end{figure}

Included within the \Chandra GC survey was the X-ray source known as
the ``bursting pulsar'', GRO~J1744-28.  This source was discovered by
BATSE in 1995 and is only the second system known to have exhibited
Type II bursts.  The accuracy of the measured position of GRO~J1744-28
has improved dramatically over the time since its discovery.  The
current best measurement of its position is that of Wijnands \& Wang
(2002) obtained with \Chandra.  With the VLT we detected one stellar
source within the \Chandra error circle (Figure 6).  Its position is
$R.A. = 17^{\rm h}44^{\rm m}33^{\rm s}.07$, $Dec. =
-28^{\circ}44'26$\arcsec.89 with a 0\arcsec.1 error. This source was
detected in all three bands with observed magnitudes of $J = 18.98
\pm0.02$, $H = 15.68 \pm 0.01$ and $K_{S} = 14.43 \pm 0.01$ (the
photometry was calibrated using 2MASS).  The measured extinction for
the source is $A_{K} = 1.64$, $A_H = 2.80$ and $A_J = 4.86$. This
gives the prospective counterpart to GRO~J1744-28 an intrinsic
magnitude of $J = 14.12$, $H = 12.88$ and $K_{S} = 12.79$. With these
magnitudes and colours, the star is likely to be an M giant.

\begin{figure}[!t]
\centering{
  \includegraphics[width=0.6\columnwidth]{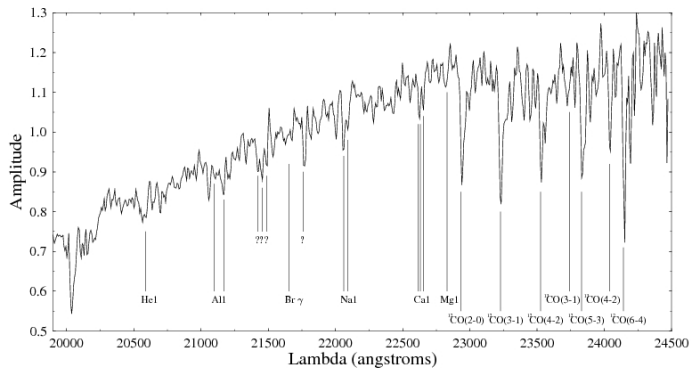}}
\vspace{-0.5cm}
  \caption{$K$-band spectrum of the candidate counterpart to the
  bursting pulsar (GRO~J1744-28).  The strong CO bands indicate that
  this star is a late-type giant.  The lack of Br $\gamma$ emission
  suggests that this star is likely not the true counterpart to the
  X-ray pulsar.}
\end{figure}

As part of our spectroscopic program, we obtained a $K$-band spectrum
of the candidate counterpart to GRO~J1744-28 (Figure 7).  From the
absorption lines in the spectra, it appears that the potential
counterpart is indeed an M type giant. There are strong $^{12}$CO
bandheads, some evidence of $^{13}$CO bandheads, and Mg, Ca and Na
absorption features. The presence of an Fe line near the
$^{13}$CO(3-1) bandhead is a feature found in M2+ III giants.  No
emission lines are evident.

Taking values for a typical NS and a late M III giant produces results
for the dynamical and physical properties of this binary system that
suggest this star is unlikely to be the true counterpart.  A typical
late M III has $M_{\rm c} = 1.4 \pm 0.4\, {\rm M_{\odot}}$ and $R_{\rm
c} = 131.9 \pm 11.7\, {\rm R_{\odot}}$.  Taking the extremes of NS
masses from the literature, we use a typical NS of $M_{\rm x} = 1.6
\pm 0.6\, {\rm M_{\odot}}$. Using the orbital period given by Finger
\etal (1996) of $P_{{\rm orb}} = 11.8337 \pm 0.0013\, {\rm days}$ we
find that the system has a semi-major axis of $a = 32 \pm 4\, {\rm
R_{\odot}}$ and that the Roche lobe (RL) of the giant is $R_{\rm RL} =
11 \pm 4\, {\rm R_{\odot}}$, in reasonable agreement with the values
given by Finger \etal.

It therefore seems unlikely that this star is the true counterpart as
if it was, the X-ray source would be orbiting within the envelope of
the companion. It is most likely that the source observed is simply a
coincidental astrometric match to the X-ray error circle, and that the
true counterpart is a fainter star beyond the limits of this
survey. This is highly likely due to the very high stellar density
towards the GC, where the average stellar separation in the $K_{S}$
band is $1.94$\arcsec (Gosling \etal 2006).

Our survey extended to a limiting magnitude of $K_{\rm S} = 20$.
Taking into account the measured extinction of $A_J = 4.86$, this
means that all stars of type A main sequence (MS) or earlier were
detected. Therefore, if the counterpart is a MS star beyond the limits
of this survey, we can place a limiting mass of $1.6 {\rm M_{\odot}}$
to the companion for a F0\,V type star or lower. This also places an
upper size limit of $\sim 1.5 {\rm R_{\odot}}$ for the companion. This
would indicate that the X-ray emission is not powered by Roche lobe
overflow, but more likely arises from low level accretion of the wind
of the companion. It will be necessary to carry out deeper
observations of the source to definitively identify the true
counterpart, and thus the mode of accretion.

\section{Conclusions}

\begin{figure}[!t]
\centering{
  \includegraphics[width=0.6\columnwidth]{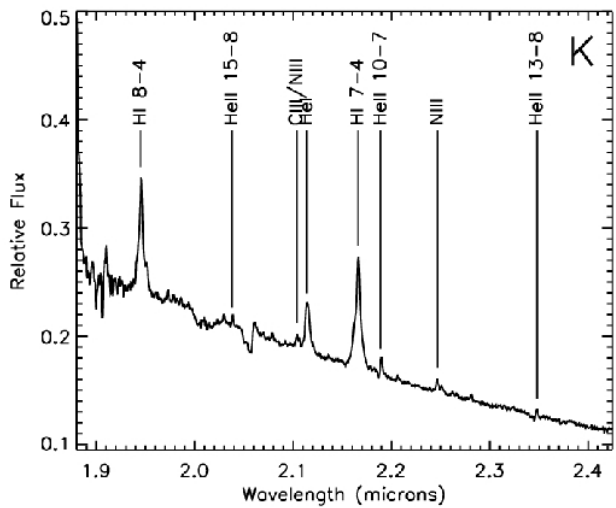}}
\vspace{-0.5cm}
  \caption{$K$-band spectrum of the first of the Galactic centre X-ray
  sources to have a spectroscopically identified counterpart.
  Emission lines that identify this star as an accreting system
  include Br $\delta$ (HI 8-4) and Br $\gamma$ (HI 7-4).}
\end{figure}

We have presented results from the search for the counterparts to the
recently discovered faint X-ray sources in the GC.  We found
astrometric matches to the positions of $\sim$70\% of our X-ray
targets.  Photometry of these prospective counterparts and follow-up
spectroscopy revealed that almost all of these candidate counterparts
were late-type K/M giant stars.  However, none of the spectra we
obtained with the VLT showed the characteristics emission signatures
expected in accreting binaries.

By combining our spectroscopic results with those of Mikles \etal
(2006), who have one confirmed spectroscopic counterpart to a \Chandra
GC source from a sample of six targets (Figure 8), we are able to draw
some preliminary conclusions about the nature of the X-ray source
population.  A small proportion, perhaps $\sim$3\%, may be HMXBs/WNS.
The remaining majority are likely to be canonical LMXBs and CVs with
late-type main-sequence mass donors with $K$ magnitudes $\geq21$ at
8.5 kpc.

\end{document}